\documentstyle[prl,aps,multicol,epsfig]{revtex} 
\draft
\title{Acoustic crystallization and heterogeneous nucleation} 
\author{X.~Chavanne, S.~Balibar and F.~Caupin}
\address{Laboratoire de Physique Statistique de l'Ecole Normale Sup\'erieure \\
 associ\'e aux Universit\'es Paris 6 et Paris 7 et au CNRS \\
 24 rue Lhomond 75231 Paris Cedex 05, France\\}
\date{2 february 2001, revised 9 april 2001}
\begin{document}
\maketitle
\begin{abstract}
By focusing a high intensity acoustic wave in liquid helium, 
we have observed the nucleation of solid helium inside the wave above a certain threshold in amplitude.
The nucleation is a stochastic phenomenon. Its probability increases continuously from 0 to 1 in 
a narrow pressure interval around $P_m$ + 4.7 bars ($P_m$ = 25.3 bars is the melting pressure
where liquid and solid helium are in equilibrium). This overpressure is larger by
 two to three orders of magnitude than what had been previously observed.
Our result strongly supports the recent suggestion by Balibar, Mizusaki and Sasaki \cite
{Balibar00} that,
in all previous experiments, solid helium nucleated on impurities.
\end{abstract}
\begin{multicols}{2}
To our knowledge, no one has ever observed that a high intensity sound wave
 travelling in a liquid can crystallize this liquid. Several experiments have shown that the negative
 pressure swings of the wave can produce cavitation, 
in other words nucleate bubbles or trigger the liquid-gas transition\cite{Balibar98}. By focusing an acoustic wave 
in liquid helium, we have found that the positive swings can also trigger the liquid-solid transition. 
It occurs if the wave amplitude
 reaches a certain threshold which we have measured. We could also estimate the size of the crystallites formed 
in the wave (15 $\mathrm{\mu m}$) and their typical growth velocity (100 $\mathrm{ms^{-1}}$, close 
to the speed of sound). We believe that it is the ability of helium crystals to grow at very high speed
 which allowed us to make this observation.
The observed threshold corresponds to a pressure 4.7 bar higher than the equilibrium pressure. This
overpressure is two to three orders of magnitude larger than what had been observed in all previous
 experiments\cite{Balibar80,Tsymbalenko82,Sasaki98,Ruutu96,Elbaum}.
Our result strongly supports the recent suggestion by Balibar, Mizusaki and Sasaki \cite
{Balibar00} that,
in all these previous experiments, solid helium nucleated on favourable impurities.

Let us first describe our experimental method. 
Inside a cryostat with optical access, we have focused an acoustic wave through liquid helium 4. 
The acoustic focus is located on a clean glass plate which allows us to measure the local instantaneous 
density of liquid helium\cite{Maris} (see Fig.1). The acoustic wave is
 emitted by 
 a hemispherical piezoelectric transducer similar to the one we used for the study of cavitation
\cite{Balibar98}. 
It resonates in a thickness mode at 1.019 MHz and we pulse it with bursts of six oscillations, consequently 
about 6 $\rm \mu s$ wide. The focusing is achieved by the transducer geometry. 
Given the characteristics of our RF amplifier, we can reach a maximum acoustic power of
 $\rm  5\,\mathrm{kW.cm^{-2}}$ (200 dB) at the focus.
The transducer is gently pressed against the glass plate. At the melting pressure,
 the sound velocity is 366.3 $\mathrm{ms^{-1}}$ in the liquid phase\cite{Abraham} ,
 so that the acoustic  wavelength is
360 $\mathrm{\mu m}$. A lens of focal length 2 cm is located inside the experimental cell and focuses
an $\rm{Ar^+}$ laser beam onto the acoustic focal region. The waist 
of the laser beam at the focus is 30 $\mathrm{\mu m}$, 
less than one tenth of the acoustic wavelength. The light spot position is carefully adjusted at the center
 of the acoustic focal region by tilting the incident beam and maximizing the modulation of the reflected light
 from the acoustic wave.
 We have measured the ratio of the ac-component to 
 the dc-component of the reflected light.
\begin{figure}[htb]
\begin{center}
\epsfig{file=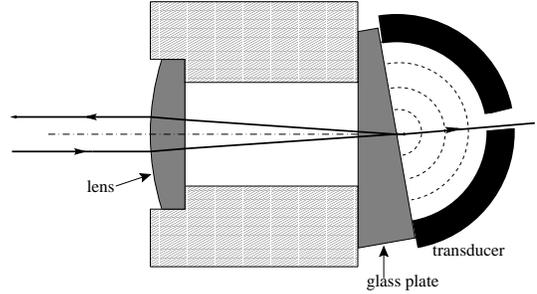,width=7cm}
\end{center}
\caption{
\label{fig:setup3}
Experimental setup}
\end{figure}
The dc-component acts as the reference. It corresponds to the reflection at the interface 
between the glass and liquid helium at the static pressure in the cell. Here we only present results 
at low temperature (T = 65 mK) when the static pressure equals $\rm P_m$.
 There is some solid helium in the bottom part of the cell. The
 transducer, the glass plate and the lens are in the liquid above. At the static pressure $\rm P_m$ = 25.324 bar,
 the static density is
$\rm \rho_m$ = 0.17245 $\mathrm{g.cm^{-3}}$. We calculated the refractive index $n_{He}$ of liquid helium from
 the Clausius-Mossoti relation 
\begin{equation}
\frac{n^2 -1}{n^2 + 2} = \frac{4\pi\rho \alpha_M }{3M} 
\end{equation}
where M = 4.0026 g, $\rho$ is the helium density and $\alpha_M$ is
 the molar polarizability.
For 514.5 nm green light, $\alpha _M$ = 0.1245 $\mathrm{cm^3mol^{-1}}$. This value is slightly larger than the zero
 frequency polarizability $\alpha_{M0}$ = 0.1233 $\mathrm{cm^3mol^{-1}}$ which is given by Harris-Lowe\cite{Harris} and
 cited by Donnelly\cite{Donnelly}. Indeed, the polarizability varies with frequency as measured by
Cuthbertson\cite{Cuthbertson} cited by M.H.Edwards\cite{MHEdwards}.

At the melting density (and for 514.5 nm), we found $ n_{He}$ = 1.0339.
Since the index of our BK7 glass plate is $ n_g = 1.5205$ for the same green light,
 the reflection coefficient is
\begin{equation}
R = \left( \frac{n_g - n_{He}}{n_g + n_{He}} \right) ^2 = 0.036288
\end{equation}
The ac-component is proportional to the modulation of the local density by the acoustic wave. It is measured
 with an avalanche photodiode (Hamamatsu APD module C5331-03) which has a sensitivity 
of 15 $\mathrm{kV.W^{-1}}$ in the range 4 kHz to 100 MHz. We use low light powers so that the APD responds linearly (the incident power in the experimental cell
 is about 500 $\rm \mu W$ and the power received by the diode is about 6 $\rm \mu W$). 
The dc-component is measured with a Si photodiode (Hamamatsu S1406). With its amplifier circuit, it
 has a sensitivity of 2.35 $\mathrm{kV.W^{-1}}$ at low frequency and works up to 400 kHz. We carefully calibrated the sensitivity ratio 
of our two diodes by modulating the Ar\+ laser beam at 200 kHz, inside the overlap region of their frequency bandwiths.
 From the known value of the reflection at $\rm P_m$ (the dc-component), we then obtained the value of the 
instantaneous density at the acoustic focus.

\vspace{4 mm}
FIG. 2 and FIG. 3 show two recordings of the instantaneous density as a function of time. These recordings were
 obtained with a LeCroy digital
 oscilloscope whose sampling rate is 1 GHz. Our transducer has a finite quality
 factor $Q = 50 \pm 5$ so that the wave amplitude increases during the six first periods. After six
 periods, the electric excitation stops and the amplitude relaxes to zero 
according to the same quality factor. The acoustic burst arrives at
 the focus 21.84 $\rm \mu$s after the beginning of the electrical excitation (time zero).

Two traces are
 superimposed on these figures. There is also a horizontal line which indicates the static liquid density
 $\rm \rho_m$ = 0.17245 $\mathrm{g.cm^{-3}}$. These traces 
are not one shot recordings, they correspond to averages on 1000 bursts repeated at a rate of 3 Hz. 
We checked that our results do not depend on the light intensity nor on the repetition rate. Fortunately, our plate
 resisted to a very large number of acoustic impacts.

The first trace corresponds to an excitation of 10.4 V, below the 
threshold for the nucleation of solid helium. The acoustic oscillation is unaltered. It shows a nearly
 sinusoidal wave with a smoothly varying envelope. The second trace corresponds 
to a slightly larger excitation (11.0 V) and shows a sharp increase of the density around 27.5 $\rm \mu s$, 
where the maximum sound amplitude is reached.
 We attribute this increase to the nucleation of solid
 helium at the acoustic focus. This nucleation occurs on the glass surface 
 since the reflectivity measurement only probes this surface and the 
 crystallization starts when the density is maximum there.
\begin{figure}[htb]
\begin{center}
\epsfig{file=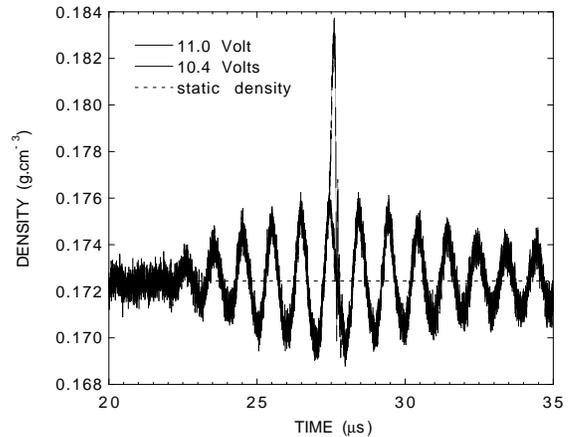,width=8cm}
\end{center}
\caption{
\label{fig:graph1}
Two recordings of the density at the acoustic focus as a function of time. One trace corresponds to 
an excitation (10.4 V) below the crystallization threshold; the other one (11.0 V) is superimposed on the first one and
 shows a peak corresponding
 to the nucleation of solid helium.}
\end{figure}
\begin{figure}[htb]
\begin{center}
\epsfig{file=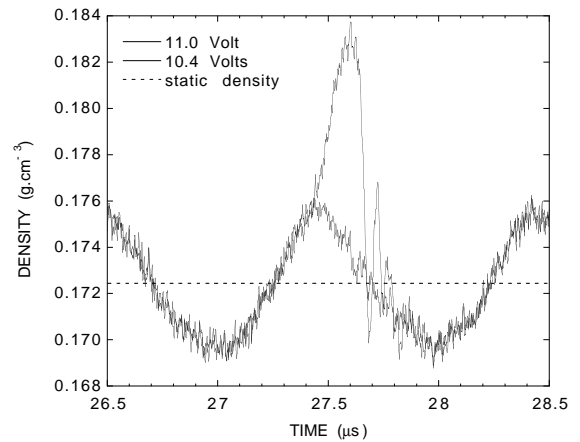,width=8cm}
\end{center}
\caption{
\label{fig:zoomgraph1}
An enlargement of the central part of Fig. 2}
\end{figure}

We have also observed that the nucleation is stochastic. The nucleation probability 
increased continuously from zero to one as the excitation increased in a small voltage interval\cite{Chavanne}. 
The probability was 3\% for 10.6 V, 35\% for 10.8 V and  98\% for 11.0 V. The first trace is 
an average on bursts which never nucleated the solid. On the contrary, the second trace is an average on 
bursts which nearly always nucleated the solid.
We also have to remember that the laser beam is focused onto a region about 30 $\mathrm{\mu m}$ wide. Since the
 density of solid helium\cite{Grilly}  is 0.19076 $\mathrm{g.cm^{-3}}$ , our observation of a maximum density
 around 0.183 $\mathrm{g.cm^{-3}}$
indicates that once the solid has nucleated, it grows up to about 15 $\mathrm{\mu m}$ in size, half the waist of
 the laser beam at the focus. As we further increased the excitation, our signal 
 saturated around 0.191 $\mathrm{g.cm^{-3}}$. It shows that the nucleated 
 phase is hcp helium.
When starting from a lower static density, we obtained the same nucleation threshold (0.1758 $\mathrm{g.cm^{-3}}$).
We expected these crystals to grow as long as the local pressure is larger than $P_m$ and 
to melt only when it becomes less than $P_m$. FIG. 3 shows that our signal decreases slightly sooner, also that
it oscillates a few times before vanishing. This surprising behaviour may be due to oscillations of the crystal
 itself (acoustic resonances inside it ?) or to the
 existence of a non-monotonic density field in the liquid.
 We also remark that the crystal grows in about 150 $\mathrm{ns}$, which corresponds 
to a very large growth velocity, about 100 $\mathrm{m.s^{-1}}$. It is well known \cite{Balibar94} that the helium crystals 
grow very fast at low temperature. With any classical crystal, the necessary evacuation of the latent heat
introduces a resistance to the growth which would hardly allow growth velocities comparable to the sound
 velocity. This is probably the reason why acoustic crystallization was never 
 observed before. Let us finally notice that the wave amplitude corresponds 
 to a relative density oscillation of order 2\%, so that the temperature at 
 the acoustic focus should also oscillate, but only by the same 2 \% if we 
 assume that the wave is adiabatic, which is reasonable.
\vspace{4 mm}

How can one understand our results? 
Several authors\cite{Balibar80,Tsymbalenko82,Sasaki98,Ruutu96,Elbaum} had observed that solid
 helium appeared in their cells if liquid
 helium was pressurized a few millibars only above the
 equilibrium pressure $P_m$.
 This is a very small overpressure. Indeed, a rough estimate\cite{Balibar00} of the energy barrier for
 homogeneous nucleation is $10^{10}$ K for an overpressure $\delta P$ = 3 mbar. Since
 Ruutu et al.\cite{Ruutu96} measured an energy barrier of order 10 K, 9 orders of magnitude less,
 Balibar et al.\cite{Balibar00} had suggested that their energy represented the pinning energy
 of an already existing liquid-solid meniscus, not the energy cost for creating the interface
 (as assumed in the elementary homogeneous nucleation theory). Balibar et al.\cite{Balibar00}
 further suggested that impurities such as graphite particles were present in all ordinary cells. 
Indeed, solid hcp helium 4 is known to grow by epitaxy on graphite
\cite{Eckstein,Balibar80,Wiechert,Maynard,Agnolet} .
If there is  a small area of such graphite somewhere in the cell, solid helium
 only needs to escape from it to invade the cell. The pinning energy being typically the product of a 
surface energy by the size of the pinning site which can be very small, Balibar et al.\cite{Balibar00} 
explained why the energy barrier for the appearance of solid helium in an ordinary cell can be
 small even for slight overpressures.
One important check of the above suggestions was to measure the nucleation threshold $\delta P$ for solid helium
 in the absence of favourable impurities. Balibar et al.\cite{Balibar00} predicted that $\delta P$ 
 would be much higher in this clean case. According to the preliminary results which are presented here,
this is indeed true.
We studied our nucleation events as a function of static pressure and temperature.
 We also performed other experiments\cite{Chavanne}
 with static pressures smaller than $P_m$ and temperatures up to 1 K. The results are similar but this study
 is not yet completed.
In this letter, we restrict our main comments to FIG.2 and FIG.3. 
These figures demonstrate that the nucleation mechanism is significantly different from what had been
 observed in all previous studies\cite{Balibar80,Tsymbalenko82,Sasaki98,Ruutu96,Elbaum}.

Indeed, 
we see that the liquid density needs to reach values as high 
as 0.1758 $\mathrm{\pm \:0.0002\: g.cm^{-3}}$ for the solid to appear. The 
main source of uncertainty is the noise in the intensity of the reflected 
light. By measuring the sound velocity as a function of pressure
 Abraham et al.
\cite{Abraham} established the following equation of state for liquid helium:
\begin{equation}
P - P_m = 567.42\:\delta \rho  + 11115\:(\delta \rho)^2 + 74271\:(\delta \rho)^3
\end{equation}
 where $\delta \rho = (\rho - \rho_m)$, the pressure P is in bars and the density $\rho$ in $\mathrm{g.cm^{-3}}$,
 $P_m$ = 25.324 bar
 and $\rho_m$ = 0.17245 $\mathrm{g.cm^{-3}}$. We note that this equation of 
 state is very accurate and almost indistinguishable from the equation
 by Maris once one corrects it for the fact that he took pressures in bars when they are
 in atm in the original paper by Abraham {\it et al}. We also note that the liquid density at melting is very slightly 
different from Grilly's value\cite{Grilly} 0.17293.
Using Abraham's measurement we deduce the value of the liquid pressure at the nucleation threshold,
 $\mathrm{P = 30.0 \pm 0.3\: bar =  P_m + 4.7 \pm 0.3\: bar}$.
We further remark that crystal seeds do not survive on the glass plate, although the static pressure differs from $\rm P_m$
 by a very small hydrostatic pressure only.
We believe that there is no defect on the glass plate which is able to keep a crystal seed, but we cannot yet exclude 
that seeds are washed out by the negative pressure swing which follows the positive one.

From the temperature variation of the nucleation statistics, we hope to measure the activation energy E. 
 It should be of order 10 K \cite{Balibar00}. Before we do this, it is interesting to estimate
 what would be the energy barrier for the homogeneous nucleation of solid helium at 
$P_m + \delta P$ = 30.0 bar, using 
 the thin wall approximation as a crude approximation\cite{Landau}. It is given by
\begin{equation}
E = \frac{16\pi}{3}\alpha^3\left(\frac{\rho_L}{(\rho_C - \rho_L)\delta P}\right)^2
\end{equation}
where $\rm \alpha = 0.17\,\mathrm{erg.cm^{-2}}$ is the average value of the surface energy of helium crystals\cite{Edwards91}. 
For $\delta P$ = 4.7 bar, we found E = 2700 K and a critical radius for the nucleation $\rm R_c = 72\: \AA$. This
 is still too large, but not at all as absurd as $10^{10}$ K. We know\cite{Balibar80,Keshishev79,Rolley95} that 
ordinary walls are partially wet by liquid helium, with a contact angle for the liquid/solid interface $\theta
 = 45 \deg$.  As explained by Uwaha\cite{Uwaha}, a critical nucleus with the shape of a truncated sphere touching the glass wall with a contact angle
$\rm \theta = 45 \deg$ would have nearly the same energy ($\rm 0.94\times  E = 2540\: K$). 

We cannot exclude that there is some
 roughness on the glass plate, which lowers the energy barrier. But the van der Waals attraction
 from the glass wall should be the most important effect. 
 One needs to account for the density gradient near this wall and to calculate the activation energy
 with  a density functional method.
Let us finally comment on homogeneous nucleation of solid helium. In this 
experiment, solid helium nucleates on the glass wall. In our previous experiments where
 the acoustic wave was focussed in the middle of liquid helium, we never saw any solid nucleation
 even by exciting our transducers with a much
 higher voltage. Given the maximum output power of our RF amplifier, we must have reached 25 + 17 = 42 bar,
 and we now know that we should have detected the solid if
 it had nucleated. At that time, we looked at the light transmitted through the focus. This is not the technique
 used for the recordings of Figs. 2 and 3, but we also looked at the transmitted light in this experiment. The 
nucleation of the solid was easy to see without any averaging, and we now use this method for the complementary measurement of 
the nucleation probability. If we use homogeneous nucleation theory, now with $\rm \delta P = 70\,\mathrm{bar}$, 
we find E = 8 K, so that a more powerful RF amplifier should allow us to observe 
the homogeneous nucleation of solid helium in the absence of glass plate. This would also be very interesting to observe.
 Indeed, at such pressures, we expect the roton gap of liquid helium to vanish, and we consider this as the spinodal limit
 for the liquid-solid transition (rotons are not related to any vortices but to local order in 
the liquid which extends to infinity if the roton gap vanishes).

\vspace{4 mm}
We have observed crystallization inside an acoustic wave. This new phenomenon 
allowed us to bring new information
 on the nucleation of solid helium, more precisely on heterogeneous nucleation
 on a clean wall, a situation which is intermediate between homogeneous nucleation and heterogeneous 
nucleation on unknown impurities. We have found that nucleation occurs 4.7 bar above the melting pressure $P_m$.
This overpressure is much more than in previous experiments where it is likeley that favourable impurites were present.
 It is also less than what we expect for homogeneous nucleation ($\rm \delta P > 70\:\mathrm{bar}$). 
 Our experiment is still in progress. We hope that it triggers new calculations of the nucleation threshold for solid helium.

We are grateful to H.J. Maris, S. Fauve and Y. Pomeau for very stimulating discussions.

\end{multicols}

\begin{references}
\bibitem{Maris} This method was originally suggested to us by H.J.Maris for the study of
cavitation. We have also observed cavitation near the glass plate. The results of this
 other experiment will be reported elsewhere.
\end{references}
\end{document}